\documentclass{aa}
\usepackage{amsmath}
\usepackage{txfonts}
\usepackage{graphicx}
\usepackage{natbib}
\usepackage{aalongtable}
\bibpunct{(}{)}{;}{a}{}{,} 

\begin{document}

\title{Search and analysis of blue straggler stars in open clusters} 
\author{
Fabrizio De Marchi \inst{1},
Francesca De Angeli \inst{2},
Giampaolo Piotto \inst{1},
Giovanni Carraro \inst{1,3,4},
\and
Melvyn B. Davies \inst{5}
}

\institute{
Dipartimento  di   Astronomia,  Universit\`a  di  Padova,
vicolo dell'Osservatorio 2, Padova, I-35122, Italy. 
\email{fabrizio.demarchi@unipd.it, piotto@pd.astro.it}
\and
Institue of Astronomy, University of Cambridge, Madingley Road,
Cambridge CB3 OHA, UK.
\email{fda@ast.cam.ac.uk}
\and
Andes Fellow, Departamento  de   Astron\'omia,  Universidad  de  
Cile,
Casilla 36-D, Santiago, Chile.
\and
Astronomy Department, Yale University, P.O. Box 208101
New Haven, CT 06520-8101 USA.
\email{gcarraro@das.uchile.cl}
\and
Lund Observatory, Box 43, SE-221 00 Lund, Sweden.
\email{mbd@astro.lu.se}
}

\date{Received / Accepted }

\abstract 
{
} 
{
This paper presents a new homogeneous catalogue of blue straggler stars (BSS) in 
Galactic open clusters.} 
{
Photometric data for 216 clusters were collected from the literature and 
2782 BSS candidates were extracted from 76 of them.}
{
We found that the anticorrelation of BSS frequency vs. total magnitude identified
in similar studies conducted on Galactic globular clusters extends to the open 
cluster regime: clusters with smaller total magnitude tend to have higher BSS frequencies. Moreover, a clear correlation between the BSS frequency (obtained normalising the total number of BSS either to the total cluster mass or, for the older clusters, to the total number of clump stars) and the age of the clusters was found. A simple model is developed here to try 
to explain this last and new result. The model allows us to ascertain the important 
effect played by mass loss in the evolution of open clusters.} 
{
}

   
\authorrunning{F. De Marchi et al.}
\titlerunning{Blue straggler stars in open clusters}   
\maketitle

\section{Introduction}

Blue straggler stars (BSS) occupy a region above the turn-off (TO)
point in the colour-magnitude diagram (CMD) of star clusters, where no 
stars are expected on the basis of standard stellar evolution, if
we assume that all cluster stars are coeval.  A number of
exotic explanations have been invoked to interpret the nature of these
stars. Briefly, it has been suggested that these stars are
the outcome of stellar collisions \citep{Ben87, Lom96} or of mass
exchange in close binary systems \citep{McC64, Egge89, Mat90}.  Other
possible scenarios include second-generation stars \citep{Egge88},
accretion of gas from the interstellar surrounding medium
\citep{Wil64}, and capture of field stars by a star cluster.

Interestingly, BSS are ubiquitous, as they have been found in
globular clusters (GCs), dwarf galaxies, open clusters (OCs), and in the
Galactic field. Therefore, important constraints on their nature and
possible formation mechanisms can be derived from a comparative study
of the BSS properties in different environments. In the past
two decades, a huge amount of new, high-quality CMDs from CCD
photometry from ground-based and space facilities have become
available. Recently, \citet[hereafter PDK04]{Pio04} published a new
photometrically homogeneous catalogue of BSS extracted from the
CMDs from HST observations of more than 1/3 of the known Galactic GCs
\citep{Pio02}. The present paper aims to complement the PDK04
catalogue by extending the study of BSS properties to the different
environment that is typical of OCs.

The catalogue presented in this paper can be considered as an important
extension of the effort 
made by \citet[hereafter AL95]{Ahu95} to
compile a catalogue of 969 BSS in 390 OCs,
based on the photometry obtained with photographic plates and
photoelectric detectors in the 60s, 70s, and 80s. The main conclusions
drawn from their catalogue are that (1) BSS are present in clusters
of all ages, (2) the fraction of BSS increases with the richness
and age of a cluster, and (3) BSS show a remarkable degree of
central concentration.  Moreover, the number of BSS over the
number of main sequence stars seems constant up to an age of 400
Myr.

The main reasons that led us to perform a new search of BSS in OCs
are the following: (1) the AL95 catalogue was based on a highly
inhomogeneous set of data, (2) a large number of new, high-quality
photometric studies of OC stars have appeared since AL95, and (3)
the BSS selection criteria needed in our opinion to be revised,
especially for young OCs. Most important, in order to estimate the
relative number of BSS in different clusters, the total number of BSS
must be normalised to some reference stellar population in the same
cluster. The choice made by AL95 to normalise the BSS counts to the
main sequence stars creates some problems, especially in young
clusters, where it is difficult to decide whether a star is a main sequence,
a blue straggler, or a field star.

The layout of this paper is as follows.  In Sect.~2 we present the
new database.  In Sect.~3 we describe a few interesting properties 
highlighted by the new catalogue. In Sect.~4 we develop
a simple model in
order to interpret the observed BSS frequency in OCs.  Section~5
presents our conclusions.

\section{The new database}

We collected a database of B, V (in some cases I) band CCD photometry 
of 216 galactic OCs from the literature. The clusters in this new catalogue have
ages ranging from $5.5$ to $10^4$ Myr. From the entire sample we 
extracted a new catalogue of 2782 BSS candidates.
This new catalogue can be considered an updated version of the AL95 catalogue.
Recent CCD photometry was collected for 136 clusters out of the original
390 in the AL95 catalogue.  Moreover, the database was extended with
80 additional OCs for which CCD photometry is now available,
but which were not included in the AL95 catalogue.

Out of the whole sample of 216 open clusters, only 76 (the oldest ones)
show BSS candidates.
Fifty-nine of these clusters had sufficiently deep photometry to be considered
suitable for the kind of analysis we were interested in, which involves estimating 
the total mass from the integrated magnitude (see Sect. \ref{integrated_magnitudes}).
Moreover, to avoid contamination effects from field stars, we 
decided to select for our analysis only those objects with a shorter distance from the 
cluster centre than the angular radius of the cluster given in the WEBDA archive \footnote{{\tt http://obswww.unige.ch/webda/navigation.html}}. 
After applying this selection, we were left with 774 BSS candidates.

The list of these 59 clusters used in our analysis and their
relevant parameters is found in Table~\ref{parameters}. Column 2 gives the
logarithm of the age, Col. 3 is the distance of the cluster in parsecs derived via isochrone fitting, Cols. 4 and 5 list the integrated magnitudes (the first value is calculated based
on the catalogue used for the present work, see Sect. \ref{integrated_magnitudes}, the second one from Lata et al., 2002),
Col. 6 is the radius in arcmin from the WEBDA archive, Cols. 7 and 8 are the observed numbers 
of BSS and clump stars inside the cluster radius respectively. Column 9 lists the photometric bands used for this work (BV or VI).

The entire catalogue (extracted from the oldest 76 clusters and without any selection
in radius) will be published 
in an electronic form at the CDS ({\tt http://cdsweb.u-strasbg.fr}), 
and available at the
Padova Globular Cluster Group web pages ({\tt http://dipastro.pd.astro.it/globulars}).

Table 2 (available on-line) lists all the 216 clusters analysed to extract the current 
catalogue. The following parameters are provided for each entry: right ascension and 
declination, reddening, distance modulus, logarithm of age, apparent cluster radius,
number of BSS candidates within the apparent radius and without limitation on the 
distance from the centre, number of clump stars,
and reference to the adopted photometry (all the references 
are listed in the file refs.dat, also available on-line).

Table 3 (available on-line) lists all the 2782 BSS candidates.
Column 1 provides the name of the cluster where the candidate was found. The presence of
an asterisk in column 2 indicates the existence of a special note referring to this entry (all the
notes are in the file notes.dat also available on-line). Cols. 3, 4, 5, and 6 give the identification number, the distance from the centre, the magnitude and the colour respectively for each candidate.

The content of the electronic tables is described in details in the file ReadMe also available on-line.

%

Although based on a smaller number of OCs than the one
by AL95, the new catalogue contains a larger number of BSS and has the advantage of being
much more homogeneous and therefore better suited to comparative
analyses of BSS in different environments.

\subsection{Selection criterion}

BSS occupy a region of the CMD usually bluer and brighter than the TO.
To determine the TO position, we superimposed the \citet{Gir00} theoretical 
isochrones to the observed CMD. The isochrone fitting also provided us
with an estimate of the apparent distance modulus and age for each cluster.

\begin{figure}[!h]
\centering
\includegraphics[width=\columnwidth]{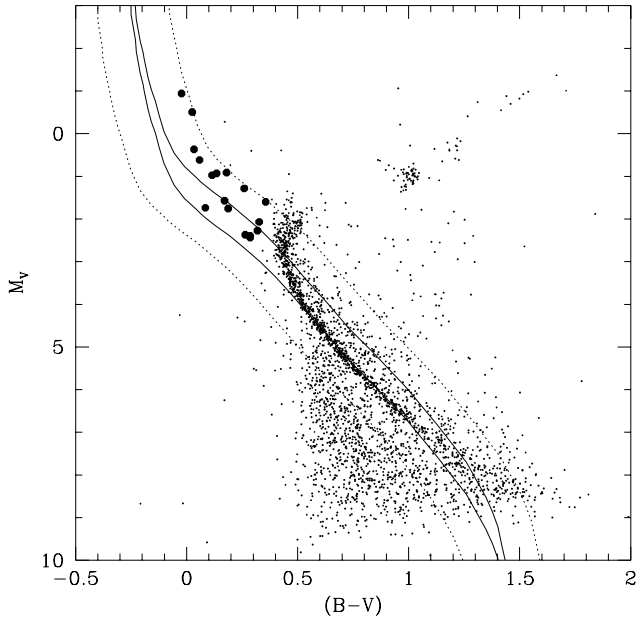}
\caption{Colour-magnitude diagram of the open cluster NGC~7789. {\em Solid lines}: 
theoretical ZAMS for single stars and for binaries with equal-mass components. 
{\em Dashed lines}: same sequences shifted by 0.15 magnitudes in colour and magnitude to take 
photometric errors and differential reddening into account. {\em Filled circles}: BSS candidates.}
\label{zams}
\end{figure}

To properly account for the photometric errors and the
differential reddening, we defined a BSS selection region in the
following way:
\begin{itemize}
\item we defined a main sequence for binaries
(${V_B}_i,\,{(B-V)_B}_i$) with equal mass components, shifting the
theoretical zero-age main sequence (ZAMS) ($V_i,\,(B-V)_i$) $0.75$ in
magnitude $V$ toward brighter magnitudes
\begin{align}
\notag
{V_B}_i & = V_i-0.75 \\
\notag
{(B-V)_B}_i & = (B-V)_i 
\end{align}

\item we then defined the two borders ({\it l} for left and {\it r} for right) of the selection region  
\begin{align}
\notag
{V_l}_i & = V_i+\Delta V \\
\notag
{(B-V)_l}_i & = (B-V)_i-\Delta (B-V) 
\end{align}
\begin{align}
\notag
{V_r}_i & = {V_B}_i-\Delta V \\
\notag
{(B-V)_r}_i & = {(B-V)_B}_i+\Delta (B-V) 
\end{align}
where $\Delta V$ and $\Delta (B-V)$ were set to 0.15 magnitudes to take 
photometric errors and differential reddening into account.
\end{itemize}
Figure \ref{zams} shows an example of the selection region on the CMD of NGC~7789.
The same procedure has been applied to the CMDs in (V, V-I).

\section{Analysis}

This section presents our analysis of the selected BSS catalogue. 
The estimate of the main parameters used in this study is explained in details and the
most relevant correlations among the observed quantities are described.

\subsection{Integrated magnitudes}
\label{integrated_magnitudes}

For each cluster we computed the integrated magnitude and the total
mass, considering as cluster members only the stars falling inside a
certain angular radius $R$ from the centre (the adopted values of $R$
are those listed on WEBDA and in Table \ref{parameters})
\footnote{Note that the lack of unresolved stars in our total magnitude
estimate is partly compensated for by the likely presence of some field
stars left in the bright star catalogue.}.

We calculated the total magnitude $M_V^{<5}$ as the sum of the luminosity of all the stars with
absolute magnitude $M_V < 5$ and angular distance from the
centre smaller than $R$. The adopted limiting absolute magnitude is a
compromise between the necessity of reaching faint magnitudes, in
order to include as many cluster stars as possible, and of having enough
clusters with deep enough available photometry to guarantee a homogeneous
determination of the total cluster magnitude.

\begin{figure}[!h]
\begin{center}
\includegraphics[width=\columnwidth]{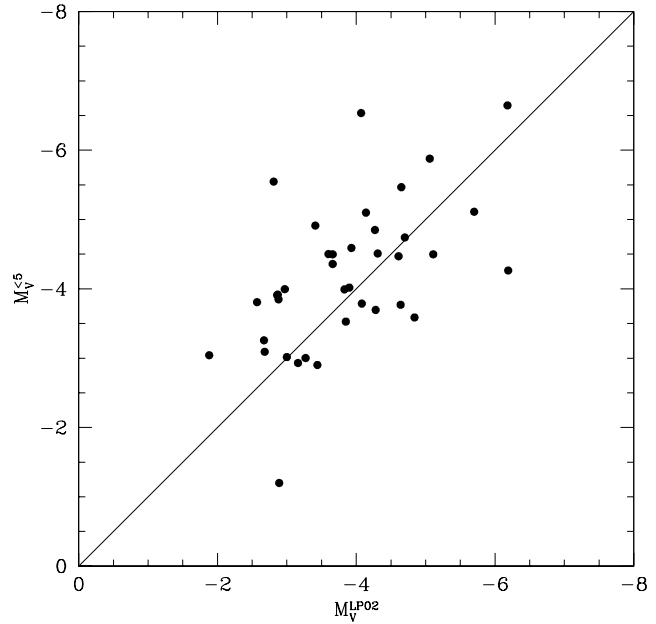}
\caption{Comparison between our integrated luminosity estimates and those 
published by LP02. The solid line merely shows the relation $M_V^{<5}=M_V^{LP02}$.}
\label{comparison} 
\end{center}
\end{figure}

Recently, \citet[hereafter LP02]{lat02} published a catalogue of
integrated magnitudes for a large sample of open clusters. Thirty-eight out of
59 clusters in our sample were listed in their catalogue.
Figure \ref{comparison} shows the comparison of LP02 values with
respect to our values for the 38 clusters in common.  There is 
overall agreement, although our integrated magnitudes are systematically
slightly brighter, probably due to contamination.

\subsection{Observed number of BSS} 

\begin{figure}
\centering
\includegraphics[width=\columnwidth]{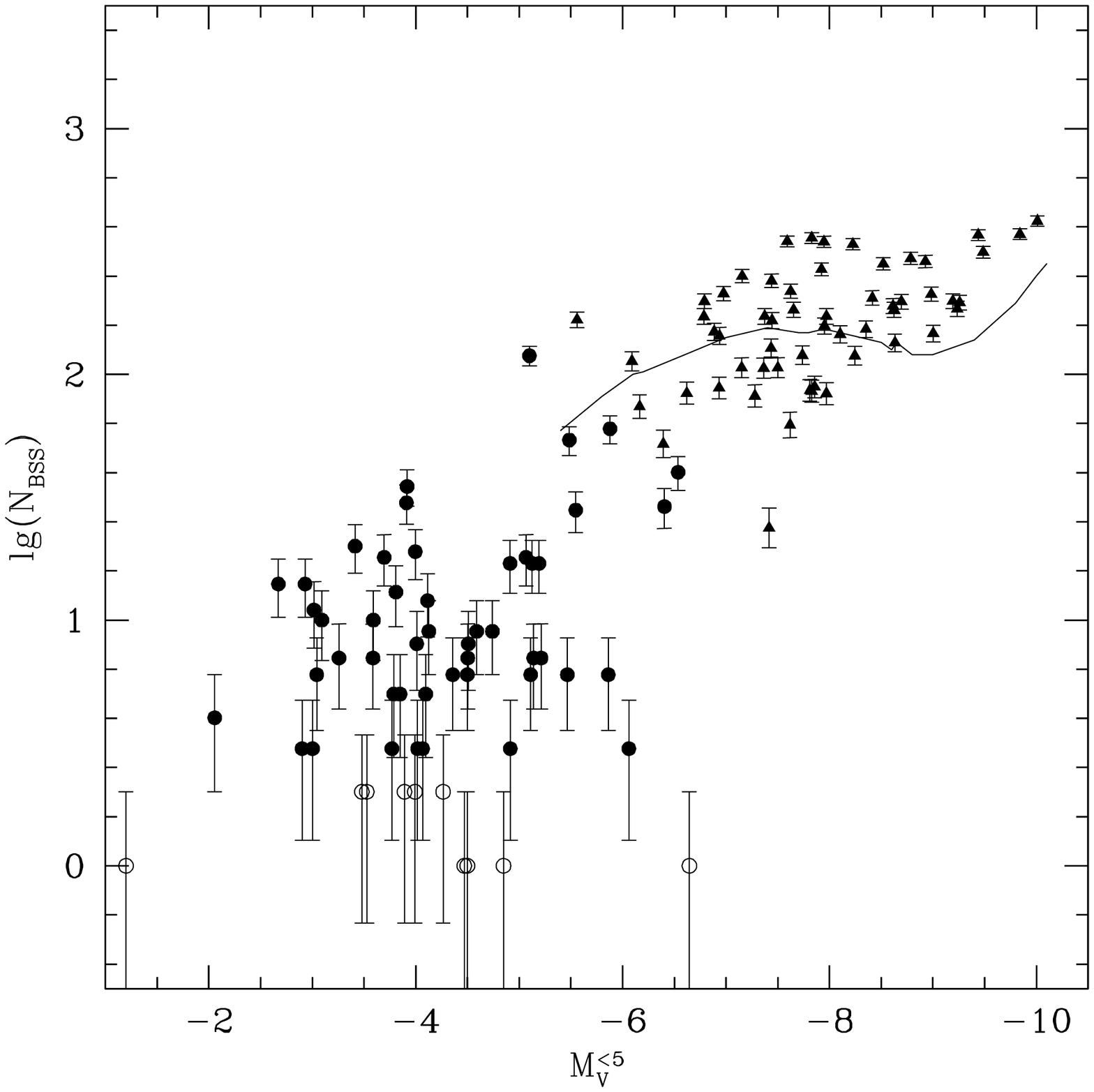}
\includegraphics[width=\columnwidth]{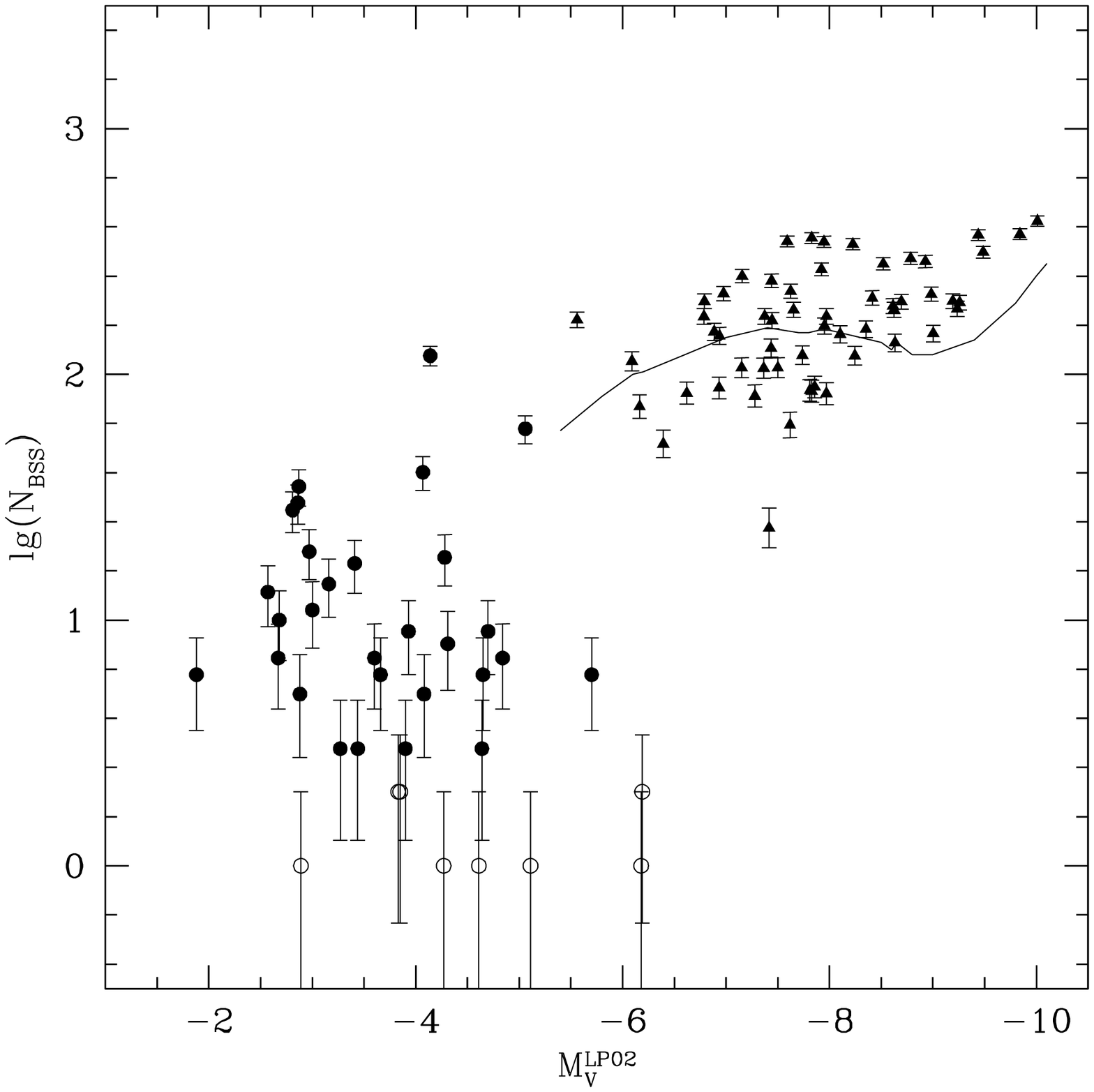}
\caption{Correlation between $\log{N_{\rm BSS}}$ and the integrated magnitude. 
Circles represent OC data, open symbols are used to mark clusters were only 1 or 2 
BSS candidate were found. The GC data from PDK04 are plotted using triangles. The
solid line shows the theoretical prediction by Davies et al., 2004.}
\label{oc-gc}
\end{figure}

The upper and lower panels of Fig. \ref{oc-gc} show the trend in the observed
number of BSS stars ($N_{\rm BSS}$) in each cluster vs $M_V^{<5}$ and 
$M_V^{\rm LP02}$, respectively. 
This figure is an extension to lower luminosity
and into the OC mass regime of Fig.~1 in \citet{Dav04}. 
The number of BSS is generally less in OCs than in GCs. 
Though with larger dispersion, the number of OC BSS versus total
magnitude seems to lie on the prolongation of the trend indicated by
the lower-luminosity GCs and by the models developed by
\citet{Dav04} (Fig. \ref{oc-gc}).
We would like to stress here that the model developed by \citet{Dav04}
is not applicable to the OC environment. It is shown here only as a useful 
comparison. 
The larger dispersion of the OC data with respect to the GC ones is due 
to the spread in age. Older clusters
have a larger number of BSS. This effect could not be observed in the GC case, as 
the spread in age is much less significative. The effects of age on the BSS 
population in OCs will be
discussed further in the next sections.

\subsection{Specific BSS frequencies}
\label{specific_frequencies}

In general, we can expect that more massive clusters contain, on
average, more stars in any evolutionary branch, including BSS.
Therefore, to properly study the BSS properties
in different clusters, it is common
practice to normalise the number of BSS to some global cluster
parameter or to the total number of stars in some well-defined region
of the CMD.

We considered two different normalisations in order to estimate the
relative frequency of BSS in each cluster:
\begin{itemize}
\item $N_{\rm BSS}/\mathcal{M}_{\rm tot}$ : number of BSS normalised to the
total mass (expressed in solar masses $\mathcal{M}_\odot$), derived from 
the total luminosity, assuming $\mathcal{M}_{\rm tot}/\mathcal{L}_{\rm tot}=
1.5\, \mathcal{M}_\odot/\mathcal{L}_\odot$
\item $N_{\rm BSS}/N_{\rm cl}$ : number of BSS normalised to the number of
clump stars (i.e., stars on the horizontal branch, when this
feature was clearly identified in the CMD).
\end{itemize}

In the following, we discuss the relevant
correlations that were found between the derived BSS frequency and other
cluster parameters.

\begin{itemize}
\item{\bf Normalisation with total mass}

Figure \ref{fbsvslogage1} shows the logarithm of the specific
frequency derived by normalising the total number of BSS to the
cluster total mass ($N_{\rm BSS}/\mathcal{M}_{\rm tot}$) plotted as a
function of the logarithm of age. There is a clear correlation between
these two quantities: older clusters have a larger relative number of
BSS. A marginal correlation between these two quantities has already been 
found by AL95; however, we believe that this is
now much more evident with the new catalogue.

PDK04 show that the relative number of BSS in GCs
anticorrelates with the cluster total luminosity, in the sense that
less luminous (less massive) GCs have a larger relative number of BSS.
Recently, this anticorrelation has been further extended to very 
low-luminosity GCs by \citet{San05}.
Figure \ref{fbsvsmagtot1} shows the specific frequency of BSS
vs the integrated magnitude. The
anticorrelation with the total magnitude continues down to $M_V^{<5}<-5$:
less massive clusters have, on average, a higher BSS frequency. In the
smallest clusters of our sample, the BSS relative frequency is about
two orders of magnitude higher than in the most massive GCs of PDK04.

\begin{figure}[!h]
\begin{center}
\includegraphics[width=\columnwidth]{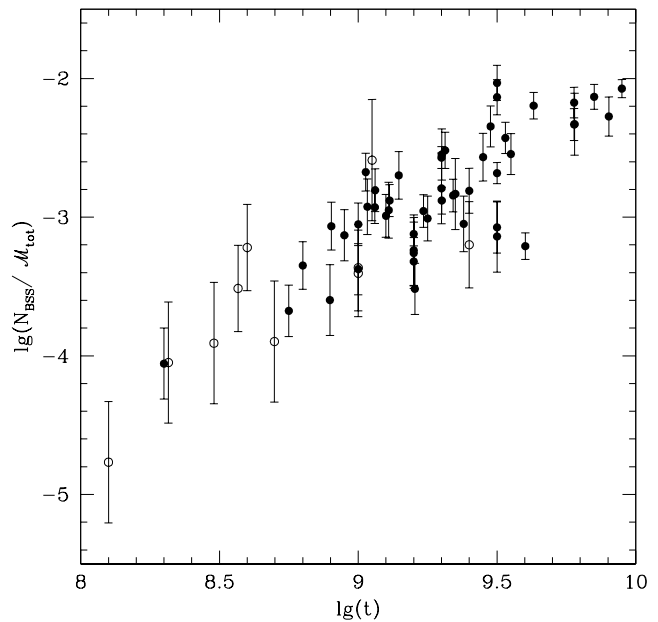}
\caption{The figure shows the trend in $\log(N_{\rm BSS}/\mathcal{M}_{\rm tot})$ 
vs the logarithm of age. Open
symbols are used to mark clusters with only 1 or 2 BSS candidates.}
\label{fbsvslogage1}
\end{center}
\end{figure}

\begin{figure}[!h]
\begin{center}
\includegraphics[width=\columnwidth]{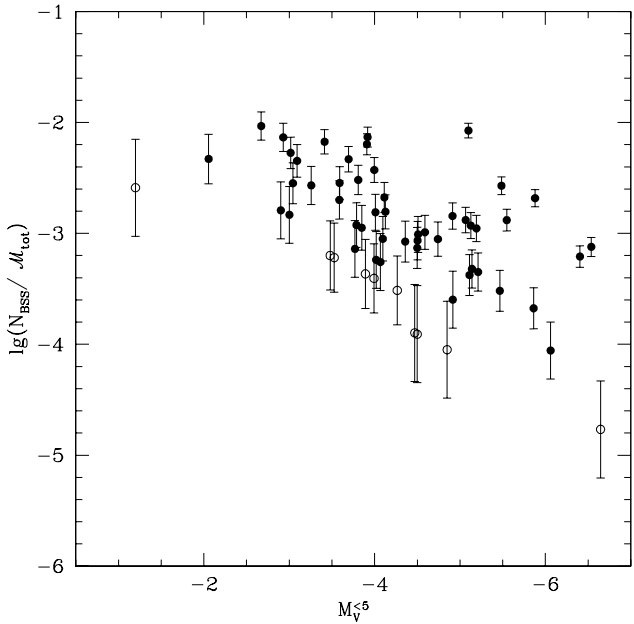}
\caption{The figure shows the trend in $\log(N_{\rm BSS}/\mathcal{M}_{\rm tot})$ 
vs the integrated magnitude. Open
symbols are used to mark clusters with only 1 or 2 BSS candidates.}
\label{fbsvsmagtot1}
\end{center}
\end{figure}

\item{\bf Normalisation with clump stars}

The selection of clump stars in our CMDs was often very uncertain. The
clump region in OCs is much less clearly defined and much
less populated than the horizontal branch region in GCs. 
Only 31 out of the 59 clusters show a clear clump
population. These are the oldest clusters in our sample (all older 
than $\sim10^{9}$ years). The number of clump stars in
every single cluster of our database is always smaller than
10. Nevertheless we do not expect this number to be contaminated by
field stars, given the very small area occupied by clump stars in OCs.

We then divided the subsample of 31 clusters into two groups, one
containing 17 clusters where the selection of clump stars was
straightforward and reliable, and another containing 14
clusters where the selection was uncertain mainly due to the small number of
clump stars.

Figure \ref{fbsvsmagtot2} shows the trend of $\log(N_{\rm BSS}/N_{\rm cl})$
vs  $M_V^{<5}$, and the GC data from PDK04, who calculated the 
BSS frequency by normalising the
total number of BSS to the total number of horizontal branch stars,
which is analogous to our normalisation to the clump stars. 
Noteworthy, and even more clearly so than in Fig.~\ref{oc-gc}, the trend 
in relative number of BSS vs integrated
magnitude already seen for the globular clusters also extends to
the OCs. Finally, \citet{Pre00} have
estimated a ratio $N_{\rm BSS}/N_{\rm HB}\sim4$ among field stars that 
agrees with the 
extrapolation of the correlation in Fig. \ref{fbsvsmagtot2} at
fainter integrated magnitudes.
\begin{figure}[!h]
\begin{center}
\includegraphics[width=\columnwidth]{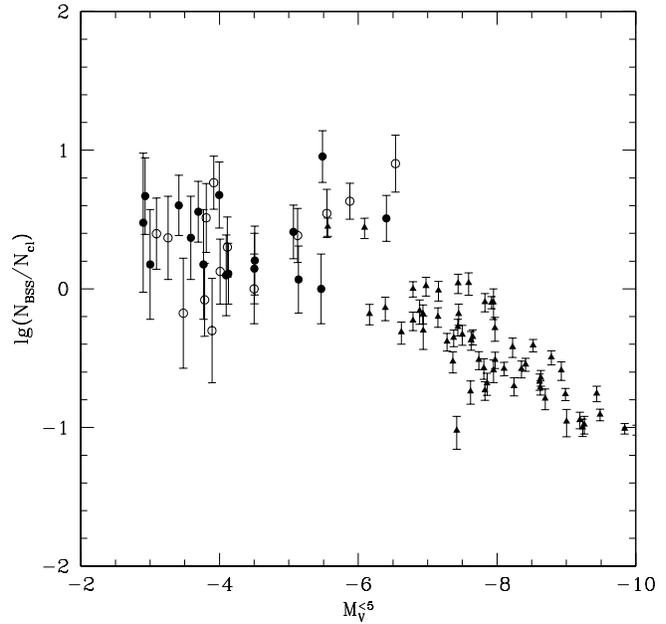}
\caption{The figure shows the trend in $\log(N_{\rm BSS}/N_{\rm cl})$ 
vs the integrated magnitude. Circles represent 
OC data and open symbols have been used to mark clusters where the selection of 
clump stars was particularly uncertain. The GC data from PDK04 are plotted using triangles.}
\label{fbsvsmagtot2}
\end{center}
\end{figure}

\end{itemize}

\section{Modelling the expected number of BSS in OCs}

\citet{Dav04} interpret the anticorrelation between the
frequency of BSS and the total cluster mass in terms of the evolution of
primordial binaries, which is affected by the stellar encounters.
However, the previous section also shows a clear correlation between
the number of BSS normalised to the total mass of the cluster and the
age of the cluster itself, with younger clusters having a lower
relative number of BSS with respect to the older ones.  The GCs in the
PDK04 catalogue were all coeval, with ages ranging from 8 to 11 Gyr
\citep{DeA05}. The OCs have a much wider spread in age, from $10^8$ to
$10^{10}$ yr. This property offers some insight into the possible
connection between cluster evolution and BSS formation.

\subsection{Model description}

In order to further investigate this result and in an attempt to
interpret it, we have developed a model following some simple
prescriptions.  We first assume that in OCs the number of BSS formed
through direct collision (dynamical BSS) is negligible compared to
the number of those formed through the evolution of primordial binaries
(primordial BSS). 
This assumption is reasonable when considering the following facts:
\begin{itemize}
\item In the first instance, one could consider the estimated number of
encounters involving single stars. The time-scale
for a single star to undergo an encounter with another star may be
approximated by \citep{Bin87}
\begin{equation}
\tau_{\rm enc} \approx 7 \times 10^{10} {\rm yr} \biggl( \frac{10^5{\rm pc}^{-3}}{n}\biggr) 
\biggl( \frac{V_\infty}{10{\rm km\, s^{-1}}}\biggr) \biggl(\frac{{\rm R}_\odot}{{\rm R}_{\rm min}}\biggr) 
\biggl( \frac{\mathcal{M}_\odot}{\mathcal{M}}\biggr),
\label{tau_enc}
\end{equation}
where $n$ is the typical number density (calculated as the total number of
stars divided by the volume in pc$^3$), and $\mathcal{M}$ the typical combined
mass of the two colliding stars (assumed to be of the order of
$2\div3 \,\mathcal{M}_\odot$). The value of $V_\infty$ can be calculated for each cluster as
$\sqrt{2 G \mathcal{M}_{\rm cluster} / R_{\rm cluster}}$.  The ratio between the age of
each cluster ($\tau_{\rm cl}$) and $\tau_{\rm enc}$
\begin{equation}
N_{\rm enc} = {\tau_{\rm cl} \over \tau_{\rm enc}}
\label{n_enc}
\end{equation}
should give a reasonable estimate of the number of encounters that
{\it could} produce BSS.  The ratio $N_{\rm enc}/\mathcal{M}_{\rm tot}$
is negligible compared to the number of observed BSS (the expected
number being between 1 and 2 orders of magnitude lower than the 
observed number of BSS). Moreover, this is likely to 
overestimate the number of collisional BSS because 1) not all the
dynamical interactions will produce BSS, 2) only BSS formed in the
past $\approx 1$ Gyr (approximate estimate of the BSS lifetime) will be
visible today.  On the other hand, this estimate does not take 
dynamical interactions involving binaries into account. These are expected
to be more frequent due to the larger cross section. However, even in
this case, we do not expect all the encounters to produce BSS.
\item The fraction of binaries undergoing a dynamical interaction can
be estimated using the former formula but adopting a value for ${\rm R}_{\rm min}$
typical of the size of a binary. This will obviously depend on the 
properties of the primordial binary population, and will be higher for low-velocity 
dispersion clusters (as the velocity dispersion decreases, wider binaries qualify
as  "hard"). 
Observational evidence \citep[for a discussion see][]{San05} shows that 
at least $\approx25\%$ of the BSS in M~67 are dynamically formed.
However, assuming a value of the order of a 100 AU for ${\rm R}_{\rm min}$ 
and using a realistic estimate of the cluster velocity dispersion in Eq. \ref{tau_enc},
one finds that the mean expected fraction of binaries experiencing dynamical 
interactions for the clusters in our sample is more than 10 times lower than
the one in M~67.
Thus we expect the fraction of dynamical BSS in most of our clusters to be 
of the order of a few percent, assuming an equal binary fraction for all clusters.
More careful considerations should also take the age of the 
clusters into account. Our sample spans an age range between $10^8$ and $10^{10} $ yr, and 
it is not at all clear how age influences the collision rate. The widest binaries 
(those with higher cross-section) are likely to interact earlier on, and their 
BSS products might not be visible today.
\end{itemize}

Our simple model builds up a cluster with both a single star and a
binary population. All the components are drawn randomly from the same
IMF \citep[we adopted the IMF of][]{Eggl89}. The masses are generated
using the equation
\begin{equation}
\mathcal{M} = \frac{0.19 x}{(1-x)^{0.75} + 0.032 (1-x)^{0.25}} \mathcal{M}_\odot,
\end{equation}
with a series of random numbers $x$.
The binary fraction is assumed to be of the order of 0.5.

Binary systems will form BSS only if one of the two following criteria
is verified:
\begin{enumerate}
\item either the mass ratio between the secondary and the primary
($q=\mathcal{M}_2/\mathcal{M}_1$) is higher than a certain $q_{\rm MT}$ (and obviously
lower than $1$); in this case the mass transfer between the primary
and the secondary is expected to be stable and to lead to the
formation of a blue straggler star with an evolved companion;
\item or the mass ratio between the secondary and the primary is
lower than $q_{\rm MT}$ but higher than a certain $q_{\rm CE}$, in
which case the mass transfer will probably be unstable and lead
to a merger of the two components of the binary system, this
time leading to the formation of a single blue straggler star.
\end{enumerate}
If $q < q_{\rm CE}$, we believe that a common envelope will form and
no BSS will originate from such systems.

The choice of the values for $q_{\rm MT}$ and $q_{\rm CE}$ was 
taken in consideration of the following facts.
Conservative mass transfer from a more-massive donor to a less-massive
receiving star is often unstable \citep{Fra02}. However, if the mass ratio
is close to unity, mass transfer will reverse the mass ratio (i.e.
the donor will quickly become the less-massive star) and mass
transfer may proceed in a stable fashion. There is evidence that this
must be the case for at least some BSS, since some field
BSS have been found in wide binaries
presumably with white dwarfs \citep{Pre00}.
As a reasonable example here, we take this critical mass ratio
to be $q_{\rm MT}=0.85$.

In systems with a mass ratio below this value, mass transfer will
be unstable, resulting in the merger of the two stars and the formation
of a blue straggler, if the mass exceeds the current turnoff. In systems with a very 
low mass-ratio, the less-massive star will be
much denser than the primary. Rather than merge, the two stars 
will form a common envelope
system where gas from the more-massive (and less dense) envelope
will smother the less-massive (and denser) star, forming a common
envelope around it and the core of the more-massive star. 
In this second case, the common envelope will be ejected as the core
and the low-mass star spiral together, and no blue straggler will
be formed. Here we assume this occurs for $q<0.4$ (i.e. $q_{\rm CE}=0.4$).

For each binary, leading to a blue straggler, three different times are defined:
\begin{enumerate}
\item ${\rm t}_{\rm ON}$ : the time at which the primary evolves off
the main sequence, and thus presumably starts the mass transfer on the
secondary or the merger between the two companions. The main sequence
lifetimes were calculated by applying the analytic formulae for stellar
evolution as a function of mass and metallicity given by
\citet{Hur00}, assuming a typical metallicity close to the solar one.
\item ${\rm t}_{\rm BSS}$ : the lifetime of the formed star. This is
calculated in a different way depending on the formation process: (1)
for mass transfers, this is the main sequence lifetime ($\tau$) of a
star of mass $\mathcal{M}_2 + 5/6\, \mathcal{M}_1$ as the expected 
transferred mass is about $5/6$ of the mass of the primary; (2) for mergers, this will have to
be calculated considering the fraction of the core mass left
available in the two merging stars. In particular, if $\mathcal{M}_1$ and $\mathcal{M}_2$
are the starting total masses of the two components, then at the time
$\tau_1$ when the primary evolves off the main sequence, we will have
that the core mass of the primary
$$\mathcal{M}_{\rm core,1} = a \mathcal{M}_1^b,$$ \citep[where $a=0.125$ and $b=1.4$,][]{vdH94} has already been transformed into helium
($\mathcal{M}_{\rm core,1}=\mathcal{M}_{He,1}$). At the same time, assuming that the
hydrogen burning proceeds linearly with time during the evolution on
the main sequence, the secondary will have burned
$$\mathcal{M}_{\rm He,2}=a \mathcal{M}_2^b \tau_1/\tau_2=\mathcal{M}_{\rm core,2} \tau_1/\tau_2.$$
The core mass of the merger product will be $\mathcal{M}_{\rm core, 1+2}=a
(\mathcal{M}_1+\mathcal{M}_2)^b$, but not all of it will be hydrogen. In fact the hydrogen
core mass will be $\mathcal{M}_{\rm H, 1+2}=\mathcal{M}_{\rm core, 1+2}-\mathcal{M}_{\rm
He,1}-\mathcal{M}_{\rm He,2}$. The main sequence lifetime will then be
$\tau_{1+2} \mathcal{M}_{\rm H, 1+2}/\mathcal{M}_{\rm core, 1+2}$.  
\item ${\rm t}_{\rm OFF}={\rm t}_{\rm ON}+{\rm t}_{\rm BSS}$ : the
expected time at which the formed star evolves off the main sequence,
leaving the CMD area where BSS are selected.
\end{enumerate}

Finally, the model loops over all the time range of interest ($10^8 <
{\rm t} <10^{10}$ Gyr) and for each time (i.e. age) counts all the BSS
having ${\rm t}_{\rm ON} < {\rm t} < {\rm t}_{\rm OFF}$.  At each time
${\rm t}$, the total mass of the cluster is estimated as the sum of
all the unevolved single stars and binaries (we consider only those
binaries in which the primary has not yet evolved off the main
sequence, $\tau_1< {\rm t}$). The total mass will then be used to
normalise the total number of BSS to allow a comparison with the
observed frequencies.

\subsection{Interpreting the empirical results}

\begin{figure}[!h]
\begin{center}
\includegraphics[width=\columnwidth]{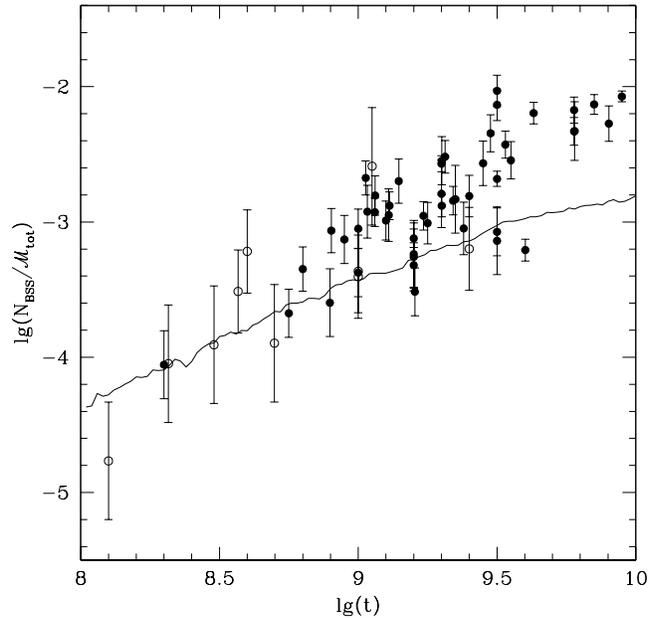}
\caption{Comparison between expected $\log(N_{\rm BSS}/\mathcal{M}_{\rm tot})$ and
observed frequencies of BSS in OC. The solid line represents the result
of our model. Observed data are shown with filled and open circles, open circles 
marking clusters with only 1 or 2 BSS candidates. }
\label{model1}
\end{center}
\end{figure}

The resulting expected time evolution of the relative number of BSS
normalised to the total mass is only slightly dependent on the assumed
values for $q_{\rm MT}$ and $q_{\rm CE}$. Changing
the values of these parameters shifts the expected trend toward higher
or lower frequencies but does not influence the
evolution rate significantly. 
Clearly many uncertainties must be taken into account when 
proceeding with this kind of analysis. Observationally, the uncertainty
on the mass-luminosity ratio and thus on the total mass clearly affects 
the level of the trend. Theoretically, the adopted values for the parameters
$q_{\rm MT}$ and $q_{\rm CE}$ can also shift the expected trend towards
higher or lower frequencies, and our values, although reasonable, have not 
been precisely determined in previous works. 
For these reasons, we concentrate in the following more on the 
general trend of the correlation than on the level of the relative numbers 
of BSS.
Figure \ref{model1} shows the expected trend 
assuming $q_{\rm MT} = 0.85$ and $q_{\rm CE}=0.4$,
with the corresponding observed
quantities. 
As shown in the plot, the expected trend of BSS frequency with age is
flatter than the observed one. Older OCs show a higher number of BSS
per unit mass than expected.

However, one important aspect has not been taken into consideration by
our model. Mass loss plays a very important role in OCs, so we should
account for it when defining the total normalising mass. Since they are more
massive than normal stars, BSS sink toward the centre quickly after
formation, which implies that the BSS population should be less
affected by mass loss than the normal stellar population.  At
increasing ages, OCs tend to lose more and more mass. Therefore the
normalising factor becomes smaller and smaller, steepening the
relation between normalised number of BSS and cluster age in Fig.
\ref{model1}. Mass loss also depends on the cluster's total mass,
which can partly account for the wide spread of the observed
quantities, together with the large uncertainties in the mass
estimates.

Recently, a couple of papers have addressed the issue of mass loss in
stellar clusters. \citet[hereafter TF05]{Tan05} and 
\citet[hereafter L05]{Lam05} present Nbody
simulation results, predicting mass loss timescales of star clusters.
We applied the results of both works to investigating the effect of mass
loss on our predictions.

Figure 2 in TF05 reproduces the mass loss as a function of
time. The curves from the first panel at the top and on the left are
the most appropriate for our analysis, having lower initial
concentration and higher resolution.  Time was converted from Nbody
units to Myr by adopting the typical cluster properties in 5 different
age intervals. This gave us the typical mass loss in each interval,
which we then applied to the results of our model, assuming that BSS are
not affected by mass loss.  We repeated this procedure for all the
curves in the first panel of Fig. 2 of TF05 (2k, 8k, 32k, 131k, where k
stands for 1000 particles),
\begin{figure}[!h]
\begin{center}
\includegraphics[width=\columnwidth]{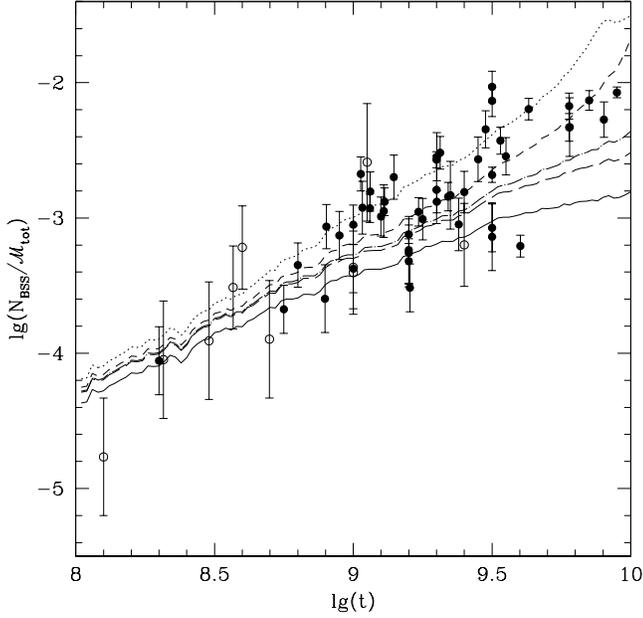}
\caption{Comparison between expected $\log(N_{\rm BSS}/\mathcal{M}_{\rm tot})$ and
observed frequencies of BSS in OC. The solid line represents the result
of our model. Observed data are shown with filled and open circles, open circles 
marking clusters with only 1 or 2 BSS candidates. Dotted,
dashed, dot-dashed and long-dashed curves reproduce respectively the
expected values obtained taking into account mass loss following the
results of Tanikawa \& Fukushige (2005) with 2k, 8k, 32k, and 131k
Nbody models.}
\label{model2}
\end{center}
\end{figure}
and Fig. \ref{model2} shows the different results. 
The estimates obtained using the 2k and 8k Nbody models are probably 
the more realistic ones, all our clusters having less 
than 2100 stars.

The observed points nicely occupy the region between the lines, 
representing the two most extreme scenarios (no mass
loss, and large mass loss affecting only normal stars), with only a
few points falling outside this region.  This may suggest that BSS are
indeed affected by mass loss as well, although not as much as normal
stars.  This preliminary analysis confirms that our simple model with
approximated mass-loss estimates can reproduce the
distribution of the observed points quite well.

To confirm these conclusions, we tried to estimate the mass
loss with a different approach. L05 provide a set of
analytical formulae to estimate the mass loss due to both stellar
evolution and tidal effects.  Stellar evolution is taken into account
by our model. Due to the
analytical form of the recipe provided by L05, it is
convenient to eliminate the tidal effects from the observed data
instead of adding them to the expected quantities.
For this reason, we use
Eq. 7 of the L05 paper to estimate the initial mass of the 
clusters in our catalogue starting from the present observed mass
(note that this equation takes into account the mass loss due to both
stellar evolution and tidal effects).
Then using Eqs. 3 and 2 of L05, we correct the calculated initial mass taking 
the mass loss due only to stellar evolution into account. 
The obtained mass estimates will be affected by mass loss due to stellar
evolution but not by mass loss due to tidal effects, i.e. will be comparable
with the masses calculated in our model (which does not account for tidal 
effects but includes stellar evolution).

\begin{figure}[!h]
\begin{center}
\includegraphics[width=\columnwidth]{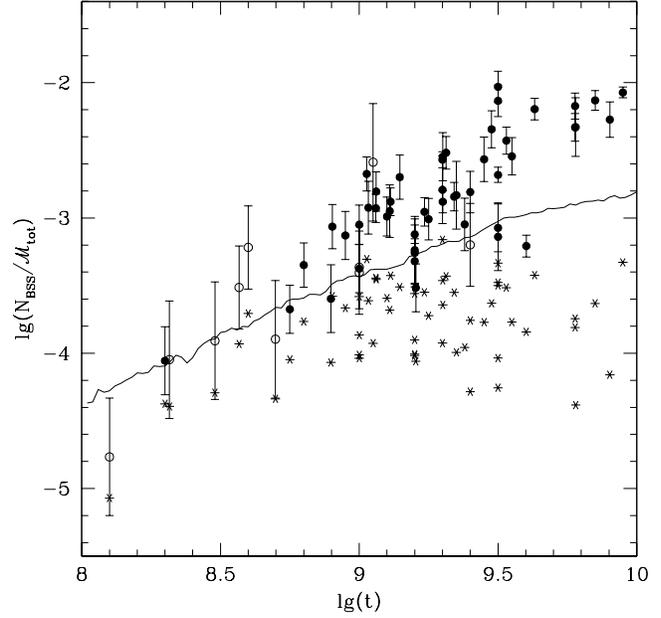}
\caption{Comparison between expected $\log(N_{\rm BSS}/\mathcal{M}_{\rm tot})$ and
observed frequencies of BSS in OC. The solid line represents the result
of our model. Observed data are shown with filled and open circles, open circles 
marking clusters with only 1 or 2 BSS candidates.
Asterisks are the observed BSS frequencies after correcting for mass loss
due to tidal effects using the recipe suggested by Lamers et
al. (2005) as explained in the text.}
\label{model3}
\end{center}
\end{figure}
Figure \ref{model3} shows the observed values, the corresponding 
values corrected for mass loss due to tidal effects applied to normal stars only 
and our model predictions.
The corrected values represent the case where the mass loss is the most evident. If 
BSS were also affected by mass loss, then the observed frequency would lie 
between the observed values and the corrected ones.
The trend predicted by the model falls in between the observed quantities and those corrected 
for the maximum tidal effect. 
Clearly the results obtained with both approaches are consistent 
with each other thus suggesting that tidal effects play an important role in OC and that 
they probably affect the number of observed BSS, as well as the total mass. 

In this respect we wish to make one last consideration. 
The fraction of binaries undergoing a dynamical interaction can be estimated 
using Eq. \ref{tau_enc} as explained in Sect. 4.1. This fraction can be estimated
for each cluster in our sample. For older clusters the expected values are higher than those
obtained for the younger ones. 
All but one of our clusters have an estimate for the fraction of binaries undergoing 
a dynamical interaction lower than the one 
calculated for M~67, the mean value being about one order of magnitude lower.
Remembering that at least 25\% of the BSS population of M~67 appears to be 
dynamically formed, this could imply that several of our clusters could show a non 
negligible dynamical population of BSS. This would also cause the
frequency of BSS in older clusters to be higher than the predictions of our model.

Finally, the normalisation with the total mass adopted for the observed number of 
BSS, is sensitive to the choice of the mass-luminosity ratio. 
The dependence of $\mathcal{M}_{\rm tot}/\mathcal{L}_{\rm tot}$ vs age 
is affected by both evolutionary and dynamical effects. Both stellar evolution
and mass loss due to tidal interactions are expected to cause the mass-luminosity 
ratio in old clusters to be higher than in young clusters. This would thus bring the
observed trend in better agreement with the results of our model.

As already noted in the previous section, there is a clear continuity
between open and globular clusters. This is clearly shown in Fig.
\ref{oc-gc}. \citet{Dav04} use a model similar to the one
implemented here to predict the trend of the total number of
BSS with respect to the total magnitude.

Although in this paper we concentrate on the evolution of the BSS
population with cluster age, we have also checked that the present
model is consistent with the predictions of \citet{Dav04} at
the low mass limit of our OCs. At ages of the order of the Galactic GC
ages ($\approx 10$ Gyr), 
the present model predicts $\log N_{\rm BSS} \simeq 3.2$. From the
total mass, predicted by the model for clusters with this age, we can estimate 
the total magnitude assuming a mass-to-light ratio of the order of 3, as 
in \citet{Dav04}. We obtain $M_V \approx -9$.
At this magnitude, their model predicts $\log
N_{\rm BSS} \approx 2.1$, including the effects of both primordial binary evolution and
stellar collisions. The dashed curve in Fig. 6 of \citet{Dav04}
shows the contribution to the total number of BSS from primordial
system. This curve would be a straight line if there were no
collisions; in fact, the decline for magnitudes brighter than $-7.4$ is
due to the increasing $N_{\rm enc}$. 
As discussed in \citet{Dav04}, the effect of the encounters is to
fasten the evolution of primordial binaries: the more encounters we
have the smaller the number of binaries able to evolve in a BSS
after 10 Gyr or so.  But this does not apply to the case of open
clusters. As shown above, we do not expect encounters to play an important
role in this environment, so that in this case we would expect the
dashed curve to be linearly increasing with $M_V$, reaching $\log
N_{\rm BSS} \approx 3.2$ at $M_V \approx -9$.

One further constraint must be satisfied by our model, i.e. the
observed frequency of BSS among field stars. \citet{Pre00} estimate
$s_{\rm BSS} = N_{\rm BSS}/N_{\rm HB} = 4$ in the field.  In order to compare
our results with this value, we estimated the number of
horizontal branch (HB) stars in our simulated clusters. We again adopted 
the analytic formulae for stellar evolution \citep{Hur00} to
evolve our simulated stars and to count the number of stars during the
HB phase at each time step.  Assuming then the much lower metallicity
($Z=0.0004$) typical of field stars as in \citet{Pre00}, we recomputed our
expected frequency of BSS with respect to the number of HB stars. For
ages of the order of $12\div13$ Gyr, we obtained $s_{\rm BSS} \approx 2.5\div 3$,
slightly lower but of the same order of magnitude as the results
obtained by \citet{Pre00}. The difference can be due to the different
model, stellar evolution prescriptions, and input quantities.
Obviously, when estimating the frequency of BSS in the field, mass loss
should no longer play an important role, and thus the correct
estimate should be given by our model without any correction for mass
loss.

\begin{acknowledgements}
FdM and GP acknowledge support by the Italian MIUR, under the
programme PRIN03.
The authors wish to thank the anonymous referee for useful comments
and suggestions that considerably improved the paper.
\end{acknowledgements}

\bibliographystyle{aa} 
\bibliography{biblio}

\begin{longtable}{lcccccccc}
\caption{\label{parameters}
List of open clusters and relevant parameters.  }\\
\hline\hline
ID        & $\log t$   & Dist. (pc) & $M_V^{<5}$ & $M_V^{\rm LP02}$ & $R$ & $N_{\rm BSS}$ & $N_{\rm cl}$ & Band \\
\hline
\endfirsthead
\caption{continued.}\\
\hline\hline
ID        & $\log t$   & Dist. (pc) & $M_V^{<5}$ & $M_V^{\rm LP02}$ & $R$ & $N_{\rm BSS}$ & $N_{\rm cl}$ & Band \\
\hline
\endhead
\hline
\endfoot
ArpMadore 2     &  9.40  &  13320   &  -3.48  & $\dots$ &  1.00   &  2    &   3     &  BV   \\
Berkeley 10     &  8.80  &   2294   &  -5.21  & $\dots$ &  5.00   &  7    & $\dots$ &  BV   \\
Berkeley 12     &  9.50  &   3289   &  -2.67  & $\dots$ &  2.00   &  14   & $\dots$ &  BV   \\
Berkeley 14     &  9.20  &   5495   &  -6.54  &  -4.07  &  2.50   &  40   &   5     &  BV   \\
Berkeley 20     &  9.78  &   8387   &  -2.06  & $\dots$ &  1.00   &  4    & $\dots$ &  BV   \\
Berkeley 21     &  9.34  &   5012   &  -4.91  &  -3.41  &  2.50   &  17   & $\dots$ &  BV   \\
Berkeley 22     &  9.35  &   6677   &  -3.00  &  -3.27  &  0.50   &  3    &   2     &  BV   \\
Berkeley 23     &  8.90  &   6918   &  -4.91  & $\dots$ &  2.00   &  3    & $\dots$ &  BV   \\
Berkeley 29     &  9.50  &  15180   &  -3.77  &  -4.64  &  3.00   &  3    &   2     &  BV   \\
Berkeley 30     &  8.48  &   4786   &  -4.50  &  -5.11  &  1.50   &  1    & $\dots$ &  BV   \\
Berkeley 31     &  9.31  &   7185   &  -3.81  &  -2.57  &  2.50   &  13   &   4     &  BV   \\
Berkeley 32     &  9.53  &   3496   &  -4.00  &  -2.97  &  3.00   &  19   &   4     &  BV   \\
Berkeley 33     &  8.70  &   3855   &  -4.47  &  -4.61  &  2.00   &  1    & $\dots$ &  BV   \\
Berkeley 39     &  9.78  &   3802   &  -3.69  &  -4.28  &  3.50   &  18   &   5     &  BV   \\
Berkeley 81     &  9.00  &   3006   &  -4.74  &  -4.70  &  2.50   &  9    & $\dots$ &  BV   \\
Berkeley 99     &  9.50  &   4943   &  -4.36  &  -3.66  &  5.00   &  6    & $\dots$ &  BV   \\
Collinder 74    &  9.11  &   2514   &  -3.85  &  -2.88  &  2.50   &  5    & $\dots$ &  BV   \\
Collinder 110   &  9.11  &   2415   &  -5.07  & $\dots$ &  9.00   &  18   &   7     &  BV   \\
Collinder 261   &  9.85  &   2518   &  -3.92  &  -2.87  &  4.50   &  35   &   6     &  BV   \\
IC 1311         &  9.20  &   6289   &  -4.02  &  -3.90  &  2.50   &  3    & $\dots$ &  BV   \\
IC 4651         &  9.05  &    882   &  -1.20  &  -2.89  &  5.00   &  1    & $\dots$ &  BV   \\
King 2          &  9.78  &   6003   &  -3.41  & $\dots$ &  2.50   &  20   &   5     &  BV   \\
King 11         &  9.55  &   2500   &  -3.59  & $\dots$ &  2.50   &  10   & $\dots$ &  BV   \\
Melotte 66      &  9.60  &   4018   &  -6.40  & $\dots$ &  7.00   &  29   &   9     &  VI   \\
Melotte 105     &  8.32  &   2210   &  -4.85  &  -4.27  &  2.50   &  1    & $\dots$ &  BV   \\
NGC 188         &  9.63  &   2047   &  -3.91  &  -2.86  &  8.50   &  30   & $\dots$ &  BV   \\
NGC 1193        &  9.90  &   5207   &  -3.02  &  -3.00  &  1.50   &  11   & $\dots$ &  BV   \\
NGC 1245        &  8.90  &   3012   &  -4.50  &  -3.60  &  4.50   &  7    &   5     &  BV   \\
NGC 1907        &  8.57  &   1558   &  -4.26  &  -6.19  &  3.50   &  2    & $\dots$ &  BV   \\
NGC 2112        &  9.30  &    850   &  -3.04  &  -1.88  &  9.00   &  6    & $\dots$ &  BV   \\
NGC 2141        &  9.40  &   3802   &  -4.01  & $\dots$ &  5.00   &  8    &   6     &  BV   \\
NGC 2158        &  9.30  &   3899   &  -5.55  &  -2.81  &  2.50   &  28   &   8     &  BV   \\
NGC 2194        &  8.75  &   2805   &  -5.86  & $\dots$ &  4.50   &  6    & $\dots$ &  BV   \\
NGC 2204        &  9.20  &   3981   &  -5.47  &  -4.65  &  5.00   &  6    &   6     &  BV   \\
NGC 2243        &  9.45  &   3976   &  -3.26  &  -2.67  &  2.50   &  7    &   3     &  BV   \\
NGC 2266        &  9.00  &   3148   &  -3.99  &  -3.83  &  2.50   &  2    & $\dots$ &  BV   \\
NGC 2420        &  9.30  &   2449   &  -2.90  &  -3.44  &  2.50   &  3    &   1     &  BV   \\
NGC 2477        &  9.00  &   1259   &  -5.11  &  -5.70  &  7.50   &  6    & $\dots$ &  BV   \\
NGC 2506        &  9.25  &   3089   &  -4.51  &  -4.31  &  6.00   &  8    &   5     &  BV   \\
NGC 2627        &  9.15  &   1858   &  -3.59  &  -4.84  &  4.00   &  7    &   3     &  VI   \\
NGC 2660        &  9.03  &   2328   &  -3.79  &  -4.08  &  1.75   &  5    &   6     &  BV   \\
NGC 2682        &  9.50  &    865   &  -2.93  &  -3.16  &  12.50  &  14   &   3     &  BV   \\
NGC 5999        &  8.60  &   2046   &  -3.53  &  -3.85  &  1.50   &  2    & $\dots$ &  BV   \\
NGC 6005        &  8.95  &   2018   &  -4.50  &  -3.66  &  2.50   &  6    &   6     &  BV   \\
NGC 6067        &  8.10  &   2241   &  -6.66  &  -6.18  &  7.00   &  1    & $\dots$ &  BV   \\
NGC 6253        &  9.48  &   1727   &  -3.09  &  -2.68  &  2.00   &  10   &   4     &  BV   \\
NGC 6791        &  9.95  &   3695   &  -5.10  &  -4.14  &  5.00   &  119  & $\dots$ &  BV   \\
NGC 6819        &  9.38  &   2349   &  -4.10  & $\dots$ &  2.50   &  5    &   4     &  BV   \\
NGC 6939        &  9.06  &   1862   &  -4.13  & $\dots$ &  5.00   &  9    &   7     &  BV   \\
NGC 7044        &  9.10  &   2989   &  -4.59  &  -3.93  &  3.00   &  9    & $\dots$ &  VI   \\
NGC 7789        &  9.23  &   1888   &  -5.19  & $\dots$ &  12.50  &  17   & $\dots$ &  BV   \\
Pismis 2        &  9.06  &   3311   &  -5.13  & $\dots$ &  2.00   &  17   &   7     &  BV   \\
Pismis 3        &  9.03  &   1361   &  -4.11  & $\dots$ &  2.50   &  12   &   6     &  BV   \\
Pismis 18       &  9.00  &   1941   &  -3.89  & $\dots$ &  2.00   &  2    &   4     &  BV   \\
Saurer 2        &  9.30  &   6274   &  -5.48  & $\dots$ &  2.00   &  54   &   6     &  VI   \\
Tombaugh 1      &  9.20  &   3006   &  -4.07  & $\dots$ &  2.50   &  3    & $\dots$ &  VI   \\
Tombaugh 2      &  9.20  &   9419   &  -5.14  & $\dots$ &  1.50   &  7    &   6     &  BV   \\
Tombaugh 5      &  8.30  &   1754   &  -6.06  & $\dots$ &  7.00   &  3    & $\dots$ &  BV   \\
Trumpler 5      &  9.50  &   2995   &  -5.88  &  -5.06  &  7.00   &  60   &   14    &  BV   \\
\end{longtable}

\end{document}